\documentclass[reprint, aps, showkeys]{revtex4-2}
\usepackage{amsmath,amssymb,graphicx,enumerate,hyperref,natbib}
\usepackage[utf8]{inputenc}
\usepackage[english]{babel}
\usepackage{subfig}

\usepackage{xcolor,ulem}

\newcommand{\mathd}{\mathrm{d}} \newcommand{\mathe}{\mathrm{e}}
\newcommand{\mathi}{\mathrm{i}} \newcommand{\nocomma}{}
\newcommand{\tmop}[1]{\ensuremath{\operatorname{#1}}}
\newcommand{\tmverbatim}[1]{\text{{\ttfamily{#1}}}}

\begin{document}

\title{Monte Carlo study of the classical antiferromagnetic
  $J_1$-{\nobreak}$J_2$-{\nobreak}$J_3$ Heisenberg model on a simple
  cubic lattice}

\author{A.N. Ignatenko} \email[Email: ]{andrey-n-ignatenko@mail.ru}

\author{S.V. Streltsov} \email[Email: ]{streltsov@imp.uran.ru}

\author{V.Yu. Irkhin} \email[Email: ]{valentin.irkhin@imp.uran.ru}
\affiliation{M.N. Mikheev Institute of Metal Physics of Ural Branch of
  Russian Academy of Sciences}

\keywords{J1-J2-J3 Heisenberg model; simple cubic lattice; frustration;
Monte Carlo simulation; Tyablikov approximation}

\begin{abstract}
  An extensive Monte Carlo study of the classical Heisenberg model on
  a simple cubic lattice with antiferromagnetic exchange interactions
  $J_n$ between the first, second, and third neighbors is performed in
  a broad region of $J_2 / J_1$, $J_3 / J_1$ ratios, and
  temperature. The character of the phase transitions is analyzed via
  the Binder cumulant method. The Neel temperature $T_{\tmop{N}}$ and the
  frustration parameter (the ratio $f= |\theta|/T_{\tmop{N}}$,
  $\theta$ being the Curie-Weiss temperature) are calculated. A comparison with the Tyablikov approximation is carried out. The strength of the frustration effects is explored.
  Possible applications to antiferromagnetic perovskites, such as CaMnO$_3$ and HgMnO$_3$,
  are discussed.
\end{abstract}

\maketitle

\section{Introduction}

The thermodynamic behavior of frustrated spin systems is the classical
field of contemporary condensed matter physics \cite{Diepbook,khomskii2024}.
Generally speaking, the quantum antiferromagnetic Heisenberg model has
a complicated phase diagram including magnetically ordered and
spin-liquid states, the latter being extensively investigated in the
two-dimensional case \cite{Balents2010, PhysRevB.88.165138}. The
situation for the simple cubic lattice with frustrated exchange
interactions is also interesting and complicated. In particular,
the results on the formation of spin-liquid phase are somewhat
controversial \cite{PhysRevB.94.140408, PhysRevB.95.014427}.
  
There are a number of experimental examples of related real materials such as perovskites, in particular antiferromagnetic perovskites {\textit A}MnO$_3$, where Mn ions form a simple cubic lattice, {\textit A} is a simple metal, e.g., Ca or Hg. In CaMnO$_3$,
the Neel temperature $T_{\tmop{N}}$ is only 125~K, while 
absolute value of the Curie-Weiss temperature is much higher, $\theta \simeq -500$~K\cite{PhysRevB.53.14020}. In the recently synthesized phase of HgMnO$_3$, a low 
$T_{\tmop{N}}$ = 60~K was experimentally obtained, and the
paramagnetic Curie temperature is $-153$~K \cite{Zhou2020}. The exchange parameters calculated in Ref. \onlinecite{MyakotnikovPRB2024} at Hubbard's $U=3$~eV are $J_1 = 28.55$~K, $J_2 = 3.95$~K, $J_3 = -0.23$~K for CaMnO$_3$ and
$J_1 = 13.93$~K, $J_2 = 4.18$~K, $J_3 = 0.70$~K for HgMnO$_3$. The decrease in
$T_{\tmop{N}}$ seems to be caused by a moderate frustration of exchange
interactions between the nearest and next-nearest neighbors ---
competition of magnetic orders with wave vectors ($\pi, \pi, \pi$) and
(0, $\pi, \pi$).

 From the pure theoretical point of view, the
Heisenberg model with frustrated exchange interactions 
provides a unique example of the spin model where collinear
configurations dominate quasi-classically for arbitrary ratios of the
antiferromagnetic exchange parameters. This makes it an excellent
playground for testing various theoretical and numerical approaches such
as  functional renormalization group in spin \cite{PhysRevB.106.174412} and pseudofermion 
\cite{PhysRevB.94.140408} versions, nonlinear spin
wave theory \cite{HuWang2019, VYuIrkhin1992, MajumdarDatta},
series expansion \cite{PhysRevB.95.014427}, quantum \cite{PhysRevLett.80.5196, PhysRevB.94.140408} and classical 
\cite{Pinettes1998, doi:10.1142/S0217984911026632} Monte Carlo simulations. In spite of this, no extensive Monte Carlo simulation for general ratios of exchange parameters was performed to the best of our knowledge. Here we perform such calculations taking into account  
exchange interactions up to the third neighbors in the classical model, 
which does not suffer from the sign problem for $J_2\ne 0$ as the quantum
model does \cite{PhysRevB.94.140408}. Besides that, it provides more
clear and simple picture due to absence of the spin-liquid phase.

\section{Model and calculation methods}
\begin{figure}[b!]
  \includegraphics[width=0.6\linewidth]{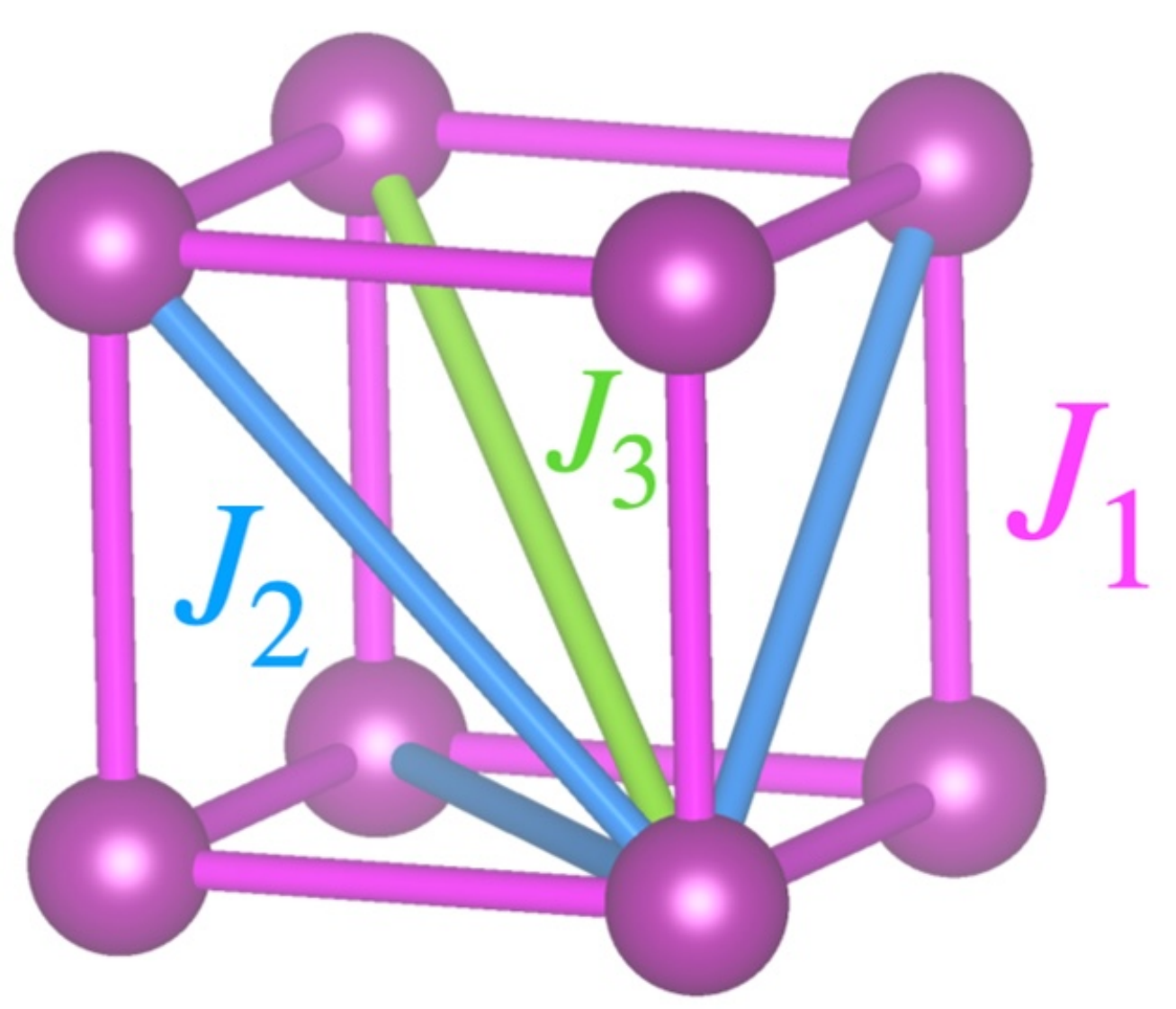} 
  \caption{Illustration of the exchange interaction parameters ($J_1, J_2, J_3$) used in the model calculations.} \label{model} 
\end{figure}

\begin{figure*}[t!]
  \subfloat{ \includegraphics[width=.33\linewidth]{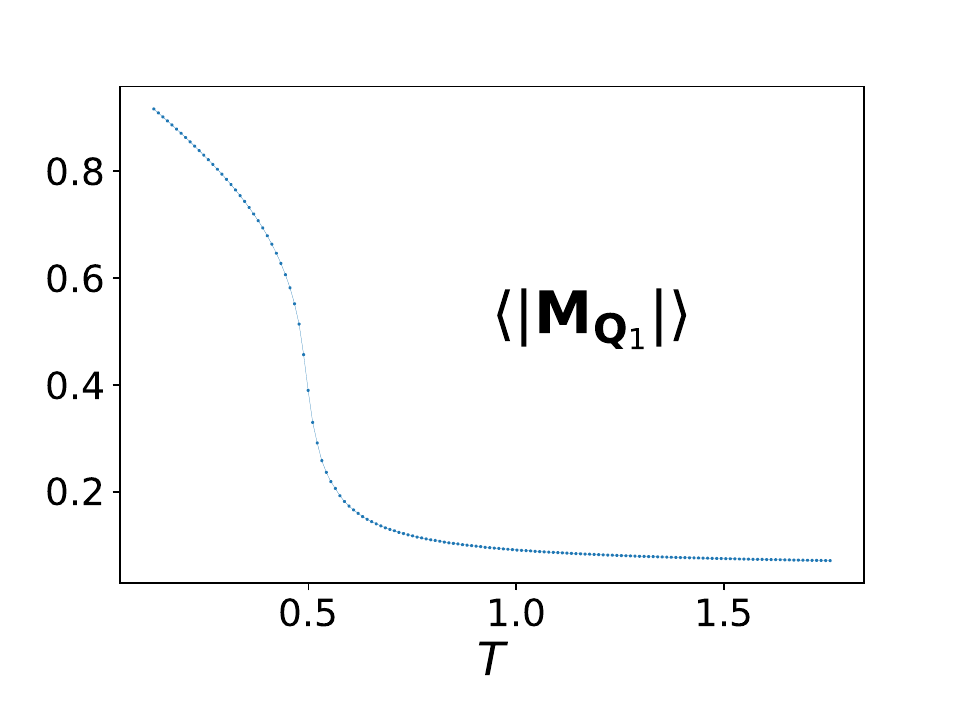} }
  \subfloat{ \includegraphics[width=.33\linewidth]{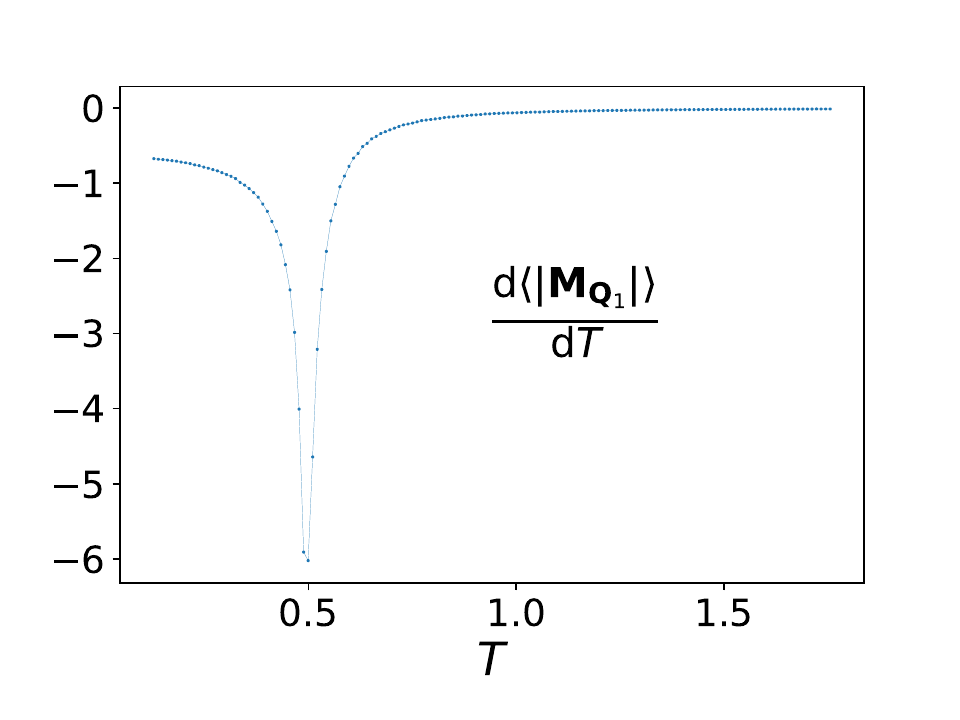} }
  \subfloat{ \includegraphics[width=.33\linewidth]{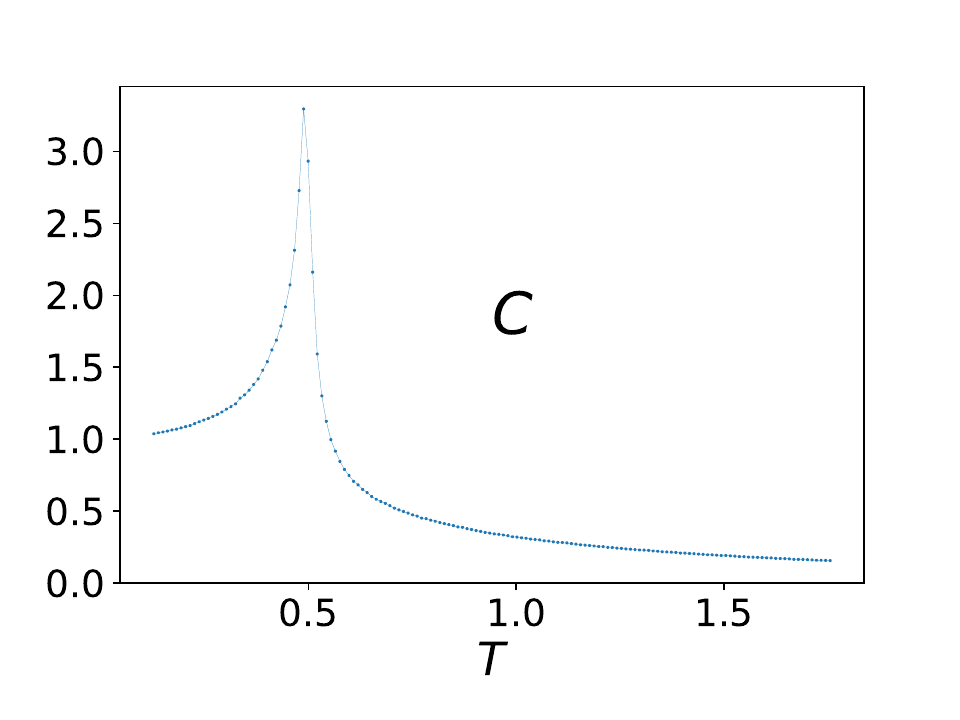}}
  \caption{Average absolute value of the total $\mathbf{Q}_1$-vector magnetic moment
  $| \mathbf{M}_{\mathbf{Q}_1} |$ as a function of temperature for
    $J_2 = 0.391$, $J_3 = 0.05$ and $L = 10$ (left panel); its
    derivative on temperature (middle panel). Temperature dependence
    of the specific heat for the same parameters (right
    panel). \label{mag_and_sh} }
\end{figure*}

The Monte-Carlo method is widely used to treat magnetic systems with localized spins (see, e.g., recent works \cite{Murtazaev2020, Sharafullin2023, ZepengZhou2023, Kolesnikov2024}). Here we apply this method to the classical Heisenberg Hamiltonian on a simple cubic lattice
\begin{eqnarray}
  H & = & \frac{1}{2} \sum_{i \nocomma j} J_{i \nocomma j}
  \mathbf{S}_i \cdot \mathbf{S}_j,
  \label{heis}
\end{eqnarray}
where each bond is counted twice in the summation, spins $\mathbf{S}_i$ are unit vectors, $\mathbf{S}_i^2 = 1$ and there is exchange interaction between the first ($J_1>0$), the second ($J_2 \geqslant 0$), and the third ($J_3 \geqslant 0$) neighbors, see Fig.~\ref{model}.  Below we suppose that the exchange parameters and temperature are all
measured in units of $J_1$, so that $J_1 = 1$. All the possible
elementary classical ground states are collinear states described by
some magnetic wave vector $\mathbf{q}$. The state with staggered order
has $\mathbf{q}=\mathbf{Q}_0 = (\pi, \pi, \pi)$. For the stripe order
state, $\mathbf{q}$ equals to one of the vectors
\begin{eqnarray*}
  & & \mathbf{Q}_1^{(1)} = (0, \pi, \pi), \quad \mathbf{Q}_1^{(2)} =
  (\pi, 0, \pi), \quad \mathbf{Q}_1^{(3)} = (\pi, \pi, 0) .
\end{eqnarray*}
For the order with alternating ferromagnetic planes $\mathbf{q}$
equals to one of the vectors
\begin{eqnarray*}
  & & \mathbf{Q}_2^{(1)} = (\pi, 0, 0), \quad \mathbf{Q}_2^{(2)} = (0,
  \pi, 0), \quad \mathbf{Q}_2^{(3)} = (0, 0, \pi) .
\end{eqnarray*}
Note that there are also more complex non-collinear ground states that
comprise of a mixture of those elementary ground states that are
degenerate in energy (see, e.g., Ref. \cite{Ignatenko2016}). For
$J_3 < 1 / 4$ the $\mathbf{Q}_0$-state is stable for $J_2 < J_3 + 1 /
4$, while for higher $J_2$ the $\mathbf{Q}_1$-state becomes
stable. For $J_3 > 1 / 4$ the $\mathbf{Q}_0$-state is stable for $J_2
< 1 / 2$, and the $\mathbf{Q}_2$-state is stable for higher $J_2$. In
this paper we restrict ourselves to the case $J_3 < 1 / 4$.

Our study of the $J_1$-{\nobreak}$J_2$-{\nobreak}$J_3$ cubic lattice
Heisenberg model is based on the classical Monte-Carlo simulation.
The standard Metropolis Monte Carlo algorithm applied to the
Heisenberg model includes the following steps:
\begin{enumerate}
\item Generate a random lattice site.
  
\item For this lattice site generate a new spin direction with uniform
  probability distribution.
  
\item Accept the new state with the probability $\exp (- \beta \Delta
  E)$, where $\Delta E$ is the change of the energy between new and
  old state and $\beta$ is the inverse temperature.
\end{enumerate}
However, because of multiple cavities in the energy landscape of the
frustrated systems, the Metropolis algorithm works poorly. The
performance is especially bad at low temperatures where  the algorithm takes
exponentially large number of steps in order to
escape from a local minimum of the energy.

Because of this drawback of the Metropolis algorithm we use 
in the calculations 
the so called ``heat bath'' algorithm defined by the following
steps
\begin{enumerate}
\item Generate a random lattice site $i$.
  
\item Calculate the field $\mathbf{h}_i = - \sum_j J_{i \nocomma j}
  \mathbf{S}_j$ produced by all other spins at the lattice site $i$.
  
\item For this lattice site generate a new spin direction with the
  probability distribution $\rho (\mathbf{S}_i) \propto \exp (\beta
  \mathbf{h}_i\nobreak\cdot\nobreak\mathbf{S}_i)$ and accept it.
\end{enumerate}
Our implementation of the ``heat bath'' Monte Carlo algorithm uses
\tmverbatim{C++} libraries ALPSCore (\url{https://alpscore.org/}, see
also Ref. \onlinecite{GAENKO2017235}) and Boost
(\url{https://www.boost.org/}), the later being basically exploited for the
Mersenne Twister pseudorandom number generator. The simulations are
performed for finite cubic supercells containing $L = 10$, 20, 30,
40 number of lattice periods and $N=L^3$ number of atoms. The periodic
boundary conditions are used. To reach the statistical convergence for
fluctuating observables, such as total energy and sublattice
magnetizations, the algorithm performs $(2 - 3) \times 10^5$ complete
sweeps of the lattice, where, by definition, one lattice sweep takes
$L^3$ elementary Monte Carlo steps, so that on average each lattice
site is touched once. The calculation of Binder cumulants (see below) requires 
much larger number of sweeps $(3 - 4) \times 10^6$.

\section{Results and discussion}

\begin{figure*}[!htb]
  \includegraphics{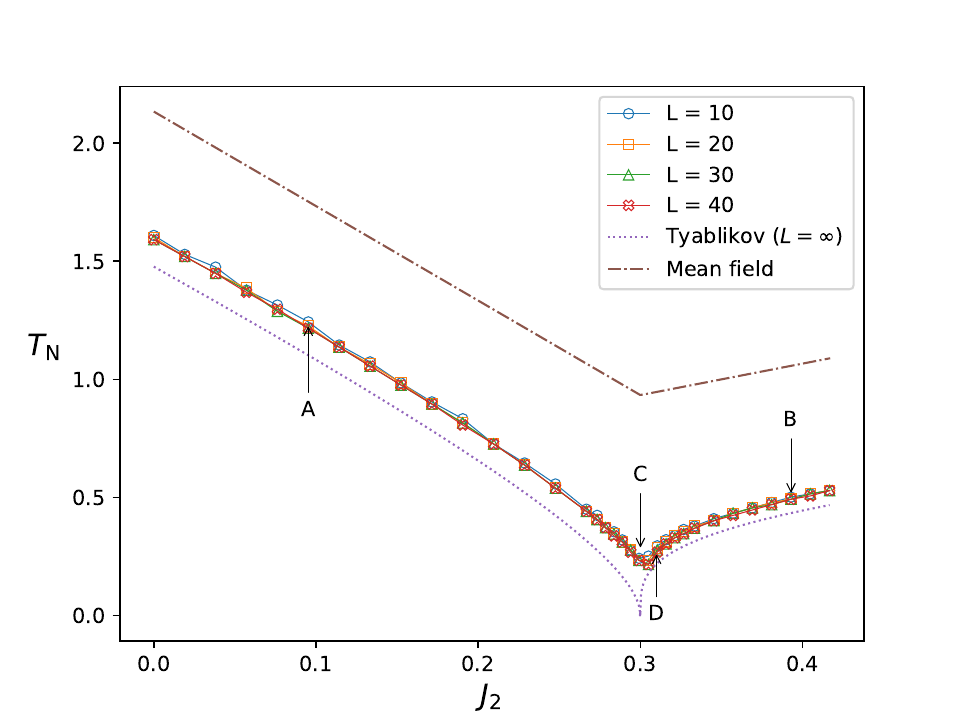}
  \caption{The Neel temperature as a function of $J_2$ calculated in
    the Monte Carlo simulation for $J_3 = 0.05$ and lattice sizes $L =
    10$, 20, 30, 40 (empty markers connected with solid lines) and compared
    with the Tyablikov approximation, $L = \infty$ (bottom dotted line)
    and the mean field theory (upper dash-dotted line).
    Arrows point to Neel temperatures for $J_2$ = 0.95 (point A),
    0.393 (point B), 0.3 = $J_2^{(\mathrm{c})}$ (point C), and
    0.31 (point D).
    \label{tn_L10-40}}
\end{figure*}

As the Monte Carlo algorithm operates,  a set of observables is
calculated and their averages are accumulated. Our set of observables
is
\begin{eqnarray*}
  & & E, \hspace{0.2em} E^2, \mathbf{M}_{\mathbf{Q}_0}, \mathbf{M}_{\mathbf{Q}_0}^2,
  |\mathbf{M}_{\mathbf{Q}_0}|, \mathbf{M}_{\mathbf{Q}_0}^4, E\mathbf{M}_{\mathbf{Q}_0}^2, \\
  & & \mathbf{M}_{\mathbf{Q}_1^{(a)}}, a = 1 \ldots 3, \quad \mathbf{M}_{\mathbf{Q}_1}^2 = \sum_{a = 1}^3
  \mathbf{M}_{\mathbf{Q}_1^{(a)}} \cdot \mathbf{M}_{\mathbf{Q}_1^{(a)}},\\
  & & 
  |\mathbf{M}_{\mathbf{Q}_1} | =  \sqrt{\mathbf{M}_{\mathbf{Q}_1}^2}, \mathbf{M}_{\mathbf{Q}_1}^4, E\mathbf{M}_{\mathbf{Q}_1}^2,\, 
\end{eqnarray*}
where $E = H / N$ is the energy per site and $\mathbf{M}_{\mathbf{q}} = N^{- 1}
\sum_i \mathbf{S}_i \mathe^{\mathi \mathbf{q}\cdot\mathbf{r}_i}$ 
is the total $\mathbf{q}$-vector magnetic moment per site. Some useful 
physical quantities can be expressed in terms of averages of these observables.
For example,
\begin{eqnarray*}
  C & = & N \beta^2 (\langle E^2 \rangle - \langle E \rangle^2)
\end{eqnarray*}
is the specific heat and
\begin{eqnarray*}
  \frac{\mathd \langle | \mathbf{M}_{\mathbf{q}} | \rangle}{\mathd T} & = &
  \beta^2 \frac{\langle \mathbf{M}_{\mathbf{q}}^2 E \rangle - \left\langle
    {\mathbf{M}_{\mathbf{q}}^2} \right\rangle \langle E \rangle}{2 \langle |
    \mathbf{M}_{\mathbf{q}} | \rangle}
\end{eqnarray*}
is the temperature derivative of the average absolute value of the total non-uniform magnetization
for a given $\mathbf{q}$-vector. This parameter tuns out to be useful in distinguishing magnetically ordered phases from the paramagnetic one. Indeed, simple
averages $\langle \mathbf{M}_{\mathbf{q}} \rangle$ or $\langle | \mathbf{M}_{\mathbf{q}}| \rangle$ cannot be used for this purpose since the first one is always zero (there is no spontaneous symmetry breaking for finite systems) and the second one is never zero. However, the temperature dependence of $\langle | \mathbf{M}_{\mathbf{q}} | \rangle$ has an inflection point (see Fig.~\ref{mag_and_sh}, left panel), corresponding to the minimum of its first derivative (see Fig.~\ref{mag_and_sh}, middle panel). The sharpness of the minimum increases with increasing system size, and the position of the minimum can be taken as an estimate for the critical temperature of the phase transition between states with $\mathbf{q}$-vector magnetic order and without such an order.

Another option is to take the position of the specific-heat maximum (see Fig.~\ref{mag_and_sh}, right panel) as a transition temperature estimate. An obvious disadvantage of this method is the inability to
determine $\mathbf{q}$-vector characterizing the magnetic order. Moreover, the maximum does not necessarily
correspond to a phase transition, for example, in two-dimensional and
layered systems the specific heat has a broad maximum at a temperature
proportional to in-layer exchange parameter~\cite{vasiliev-book}. Therefore we prefer to
determine the transition temperature from the $\langle | \mathbf{M}_{\mathbf{q}}
| \rangle$ inflection point. Nevertheless, for the present system
both the methods give almost identical critical-temperature estimates.

We compare the critical temperature determined from our Monte Carlo
simulation with the result of the mean-field approach
\begin{eqnarray*}
  T_{\tmop{N}}^{(\tmop{Mean} \tmop{field})} & = & \frac{| J
    (\mathbf{q}) |}{3},
\end{eqnarray*}
and also with  the result of the Tyablikov
approximation
\begin{eqnarray}
  T_{\tmop{N}}^{(\tmop{Tyablikov})} & = & \frac{1}{3} \left( \int
  \frac{\mathd^3 \mathbf{k}}{(2 \pi)^3} \frac{1}{J (\mathbf{k}) - J
    (\mathbf{q})} \right)^{- 1}.  \label{tyablikov}
\end{eqnarray}
In the above equations
\begin{eqnarray*}
  J (\mathbf{k}) & = & 2 (\cos{k_x} + \cos{k_y} + \cos{k_z})\\ & +
  & 4 J_2 (\cos{k_x} \cos{k_y} + \cos{k_y} \cos{k_z} + \cos{k_z}
  \cos{k_x})\\ & + & 8 J_3 \cos{k_x} \cos{k_y} \cos{k_z}
\end{eqnarray*}
(recall that $J_1 = 1$) is the Fourier transform of the exchange parameters.

Fig.~\ref{tn_L10-40} shows the dependence of the Neel temperature on
$J_2$ for $J_3 = 0.05$ and $L = 10$, 20, 30, and 40. Importantly, the
results for different values of lattice sizes $L$ are almost the same,
so it is likely that they do not change considerably in the limit $L
\rightarrow \infty$. The Neel temperature has a minimum at the point
$J_2^{(\mathrm{c})} = J_3 + 1 / 4 = 0.3$ lying at the boundary between
$\mathbf{Q}_0$ ($\tmop{lower} J_2$) and $\mathbf{Q}_1$(higher $J_2$)
ground states. Everywhere, except for the $J_2^{(\mathrm{c})}$ neighborhood,
the dependence is almost linear and has the same slope as in the
mean-field theory. As expected, the latter strongly overestimates the
Neel temperature.  The deviation from the linear dependence, being
more pronounced in the $\mathbf{Q}_1$-phase, shows the signs of the
frustration, manifested in the enhanced role of fluctuations not
accounted in the mean-field theory.

The Tyablikov approximation gives much better results and, except for
the $J_2^{(\mathrm{c})}$ neighborhood, slightly underestimates the Neel
temperature. When $J_2 \rightarrow J_2^{(\mathrm{c})}$ the integral in the Eq.
(\ref{tyablikov}) diverges and the Neel temperature vanishes, which is
not observed in the Monte Carlo simulation. This discrepancy is
explained in non-linear spin-wave theories \cite{Shender1982,PhysRevB.58.3144} (in particular, in the self-consistent spin wave theory \cite{Ignatenko2023,Ignatenko2013}). The Tyablikov
approximation, being an {\it ad hoc} version of the non-linear spin-wave theory, does not properly account for the self-energy corrections to the magnon spectrum near its hot points. Note that apparent decrease of 
the Neel temperature in the Monte Carlo simulation near $J_2^{(\mathrm{c})}$ with increasing $L$ probably is an artifact of our estimation method owing to its insufficient precision.

\begin{figure}[!htb]
  \includegraphics[width=1.1\linewidth]{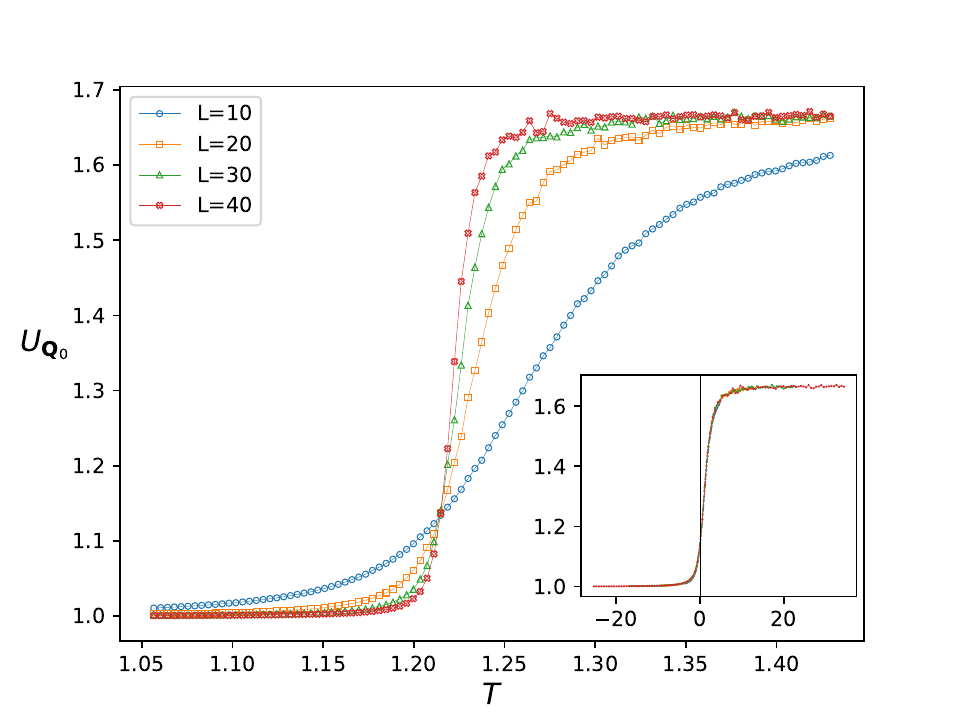}
  \caption{Temperature dependence of the Binder cumulant $U_{\mathbf{Q}_0}$
  in the vicinity of the point A ($J_2 = 0.095$, $J_3 = 0.05$) on Fig. \ref{tn_L10-40}
  for different lattice sizes $L$. $T_{\mathrm{cross}}$ = 1.2146.
  The inset shows $U_{\mathbf{Q}_0}$ as a function of $x = (T/T_{\mathrm{cross}}-1)L^{1/\nu}$ with $\nu=0.7$.
  \label{binder_a}
}
\end{figure}

\begin{figure}[!htb]
  \includegraphics[width=1.1\linewidth]{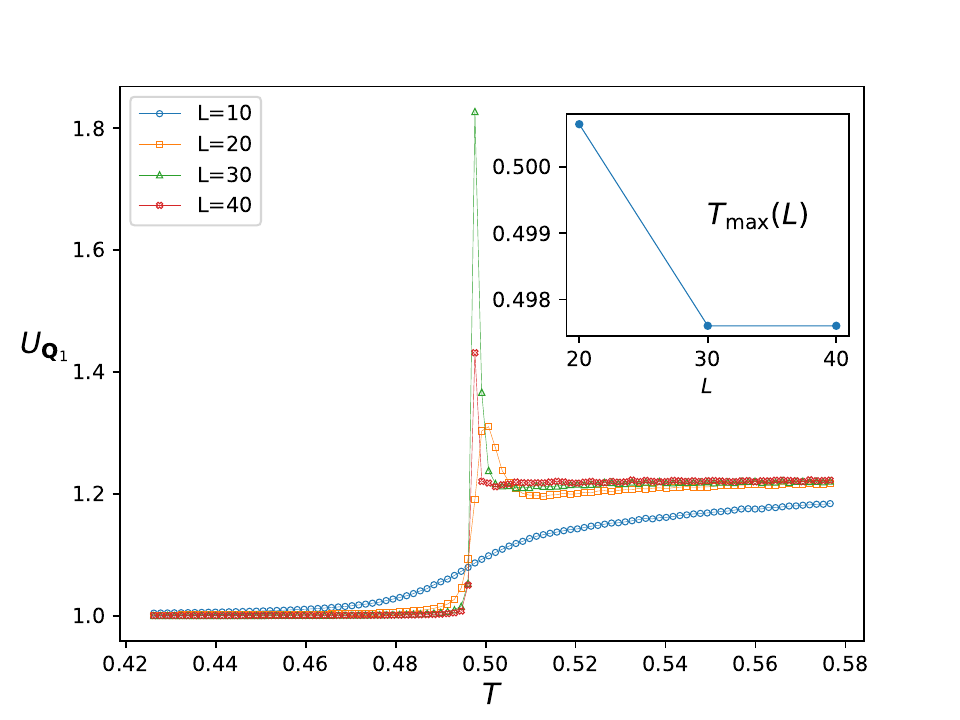}
  \caption{Temperature dependence of the Binder cumulant $U_{\mathbf{Q}_1}$
  in the vicinity of the point B ($J_2 =  0.393$, $J_3 = 0.05$) on Fig. \ref{tn_L10-40} for different lattice sizes $L$.
  The inset shows the positions of the maximums of $U_{\mathbf{Q}_1}$
  as a function of $L$.
  \label{binder_b}
  }
\end{figure}

\begin{figure}[!htb]
  \includegraphics[width=1.1\linewidth]{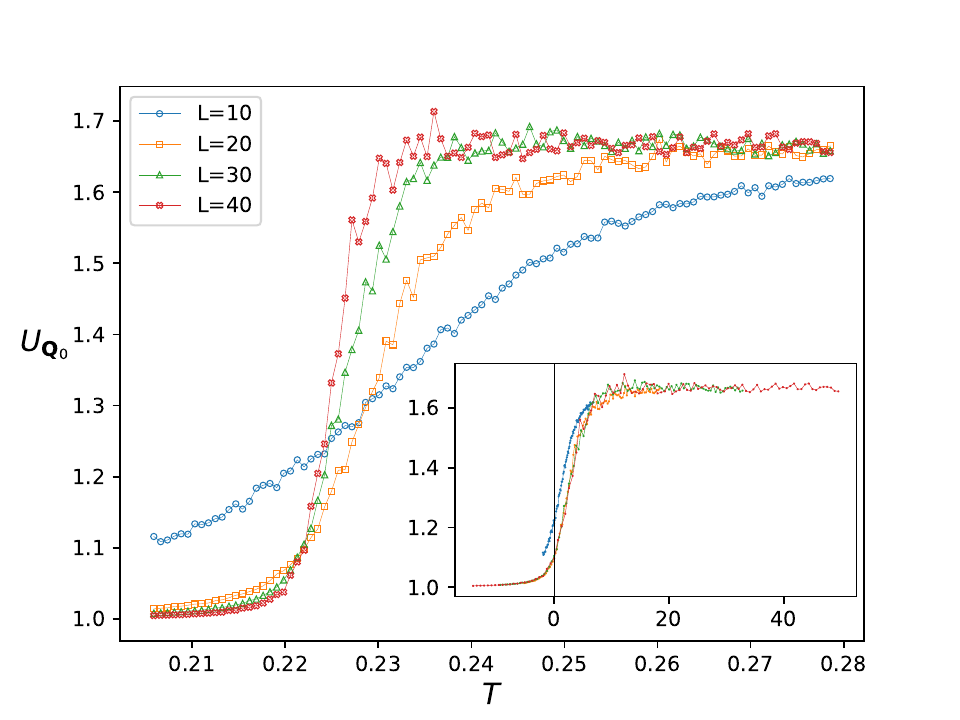}
  \caption{Temperature dependence of the Binder cumulant $U_{\mathbf{Q}_0}$
  in the vicinity of the point C ($J_2 = J_2^{(\mathrm{c})} = 0.30 $, $J_2 = 0.05$) on Fig. \ref{tn_L10-40} for different lattice sizes $L$, $T_{\mathrm{cross}}$ = 0.222. The inset shows $U_{\mathbf{Q}_0}$ as a function of $x = (T/T_{\mathrm{cross}}-1)L^{1/\nu}$ with $\nu=0.7$.
  \label{binder_c}
    }
\end{figure}

\begin{figure}[!htb]
  \includegraphics[width=1.1\linewidth]{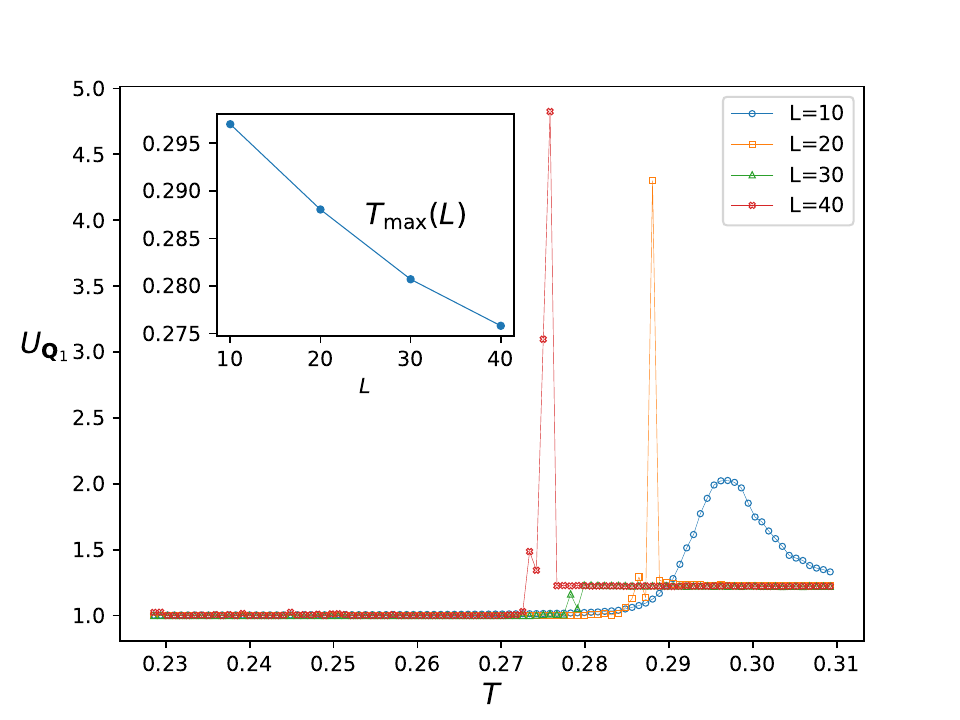}
  \caption{Temperature dependence of the Binder cumulant $U_{\mathbf{Q}_1}$
  in the vicinity of the point D ($J_2 =  0.31$, $J_3 = 0.05$) on Fig. \ref{tn_L10-40} for different lattice sizes $L$.
  The inset shows the positions of the maximums of $U_{\mathbf{Q}_1}$
  as a function of $L$.
  \label{binder_d}
  }
\end{figure}

To obtain a more precise estimate of the critical temperature and, more importantly, to determine the character of the phase transition, the Binder cumulant
\begin{eqnarray*}
U_{\mathbf{q}} = \frac{\langle\mathbf{M}_{\mathbf{q}}^4\rangle}{{\langle\mathbf{M}_{\mathbf{q}}^2\rangle}^2}
\end{eqnarray*}
can be useful \cite{Privman_book_1990}. At low temperatures, $U_{\mathbf{q}}$ is close to unity in the phase with $\mathbf{q}$-vector magnetic order. At high temperatures, when the correlation length $\xi \ll L$ and the components of the order parameter $\mathbf{M}_{\mathbf{q}}$ become almost statistically independent and fluctuate with Gaussian distribution, application of the Wick theorem gives $U_{\mathbf{q}} \approx 1 + 2/n$, where $n$ is the number of the order parameter components  ($n = 3$ for $\mathbf{q}=\mathbf{Q}_0$ and
$n = 9$ for $\mathbf{q}=\mathbf{Q}_1$).
At intermediate temperatures $U_{\mathbf{q}}$ interpolates between these two asymptotic values in a way that depends on the character of the phase transition.

In the case of a second-order phase transition, the Binder cumulant follows the scaling relation \cite{Binder_PRL_1981}
\begin{eqnarray}
U_{\mathbf{q}} = b(x = \tau L^{1/\nu}), \quad \tau = T/T_{\mathrm{N}}^{\infty} - 1,
\label{binder_scaling}
\end{eqnarray}
in the critical region $|\tau| \ll 1$ and at sufficiently large $L$, where
$T_{\mathrm{N}}^{\infty}$ is the critical temperature of the infinite system, $\nu$ is the critical exponent of the correlation length, and $b(x)$ is some function that depends on system size $L$ and temperature only via its argument $x$. In particular, the value of the Binder cumulant at $T=T_{\mathrm{N}}^{\infty}$ does not depend on $L$. Thus, one can determine $T_{\mathrm{N}}^{\infty}$ as a cross point of $U_{\mathbf{q}}(T)$ curves for two or more values of $L$.
In the case of a first-order phase transition, the Binder cumulant demonstrates a spike \cite{Binder_PRB_1984} connected with the coexistence of two phases at a transition temperature \cite{Vollmayr1993, TsaiSalinas1998}. The spike exists only when $L$ exceeds some $L_{\mathrm{min}}$. Its sharpness increases with increasing $L$ and the position drifts towards $T_{\mathrm{N}}^{\infty}$ \cite{Vollmayr1993}. 

Figs. \ref{binder_a}-\ref{binder_b} show the temperature behavior of the 
Binder cumulants $U_\mathbf{q}$ calculated in the vicinity of the points A, B, C, D 
marked by arrows in Fig. \ref{tn_L10-40}.

For the points A and C, the behavior of the Binder cumulant $U_{\mathbf{Q}_0}$ corresponds
to the picture of the second-order phase transition. Curves for different $L$ are crossed at a single point $T_{\mathrm{cross}}$, giving an estimate of $T_{\mathrm{N}}^{\infty}$. When the curves are replotted against $x = (T/T_{\mathrm{cross}}-1)L^{1/\nu}$ with $\nu=0.7$, they collapse in agreement with the scaling Eq. (\ref{binder_scaling}). Thus measured correlation-length critical exponent $\nu$ is close to $\nu_{O(3)} = 0.7112(5)$ of the three-dimensional Heisenberg universality class \cite{Campostrini_PRB_2002}, as it should be.

For the point C, the above picture is true only for $L = 20$, 30, and 40, but not for $L=10$ where the scaling seems to be broken. We can hypothesize that this means existence  of some additional characteristic length scale in Eq. (\ref{binder_scaling}), which becomes comparable with $L = 10$ at the point C and is negligibly small 
relative to $L \geqslant 20$ (a crossover scenario).

For the points B and D, the behavior of the Binder cumulant $U_{\mathbf{Q}_1}$ corresponds
to the picture of the first-order phase transition. There is no scaling, and for $L > L_{\mathrm{min}}$ we see the spikes, the sharpness of which increases with increasing $L$, and the position of the maximum converges towards some value,  providing thereby an estimate of $T_{\mathrm{N}}^{\infty}$. For the point B, $10 \leqslant L_{\mathrm{min}} < 20$, and $L_{\mathrm{min}} < 10$ for the point D. The conclusion about the first-order character of the phase transition is in agreement with the results of the $4-\varepsilon$ renormalization group treatment of a phase transition governed by the nine-dimensional order parameter \cite{Brazovskii1976}. A similar first-order phase transition was recently observed in the Monte Carlo simulation of the $J_1$-$J_2$ Heisenberg model on the body-centered cubic lattice \cite{Murtazaev2020}.

\begin{figure*}[!htb]
  \includegraphics{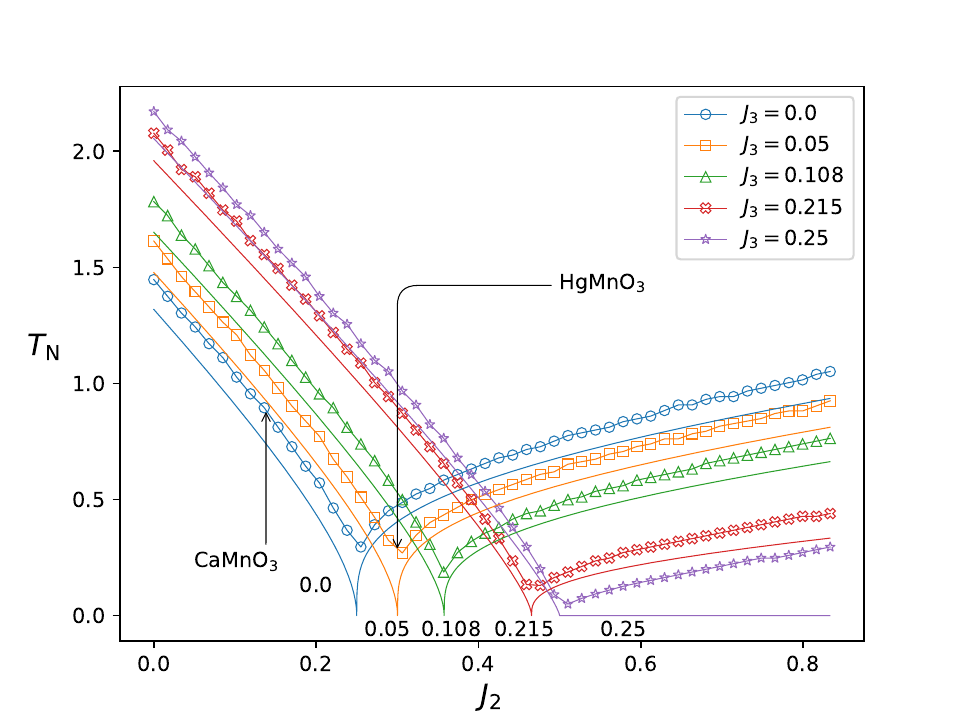}
  \caption{The Neel temperature as a function of $J_2$ calculated for
    the number of $J_3$ values in the Monte Carlo simulation for $L = 10$
    (empty markers connected with solid lines) and compared with the
    Tyablikov approximation, $L = \infty$ (solid lines with the same color as the Monte Carlo lines but without markers on them; the corresponding $J_3$-values are also indicated by the numbers near the curve's cusps). Arrows indicate Neel temperatures approximately corresponding to ratios $J_2/J_1$, $J_3/J_1$ for CaMnO$_3$ (0.138 and $-0.008$) and HgMnO$_3$ (0.3 and 0.05) taken from Ref. \onlinecite{MyakotnikovPRB2024}.
    \label{tn}}
\end{figure*}

Fig.~\ref{tn} shows the dependence of the Neel temperature on $J_2$ in
a wider interval: for $L = 10$ and for a number of $J_3$ values
increasing from $0$ to $1 / 4$. At this larger $J_2$ scale the
nonlinear character of the dependence for $J_2 > J_2^{(\mathrm{c})}$ is seen
more clearly. The value of the Neel temperature at the minimum
decreases with increasing $J_3$. At $J_3 = 1 / 4$ the Tyablikov
approximation exhibits its largest error since the integral in
Eq. (\ref{tyablikov}) diverges for all $J_2 \geqslant 1 / 2$.
The lowest Neel temperature is reached in the vicinity of the point
$(1/2, 1/4)$ in $J_2$-$J_3$ plane. At this point the Heisenberg model
(\ref{heis}) can be rewritten as a sum of complete squares of the
total spins of the cubic-lattice elementary cells, providing a higher
degree of the energy degeneracy in the ground state \cite{Balla_2019}.

\begin{figure*}[!htb]
  \includegraphics{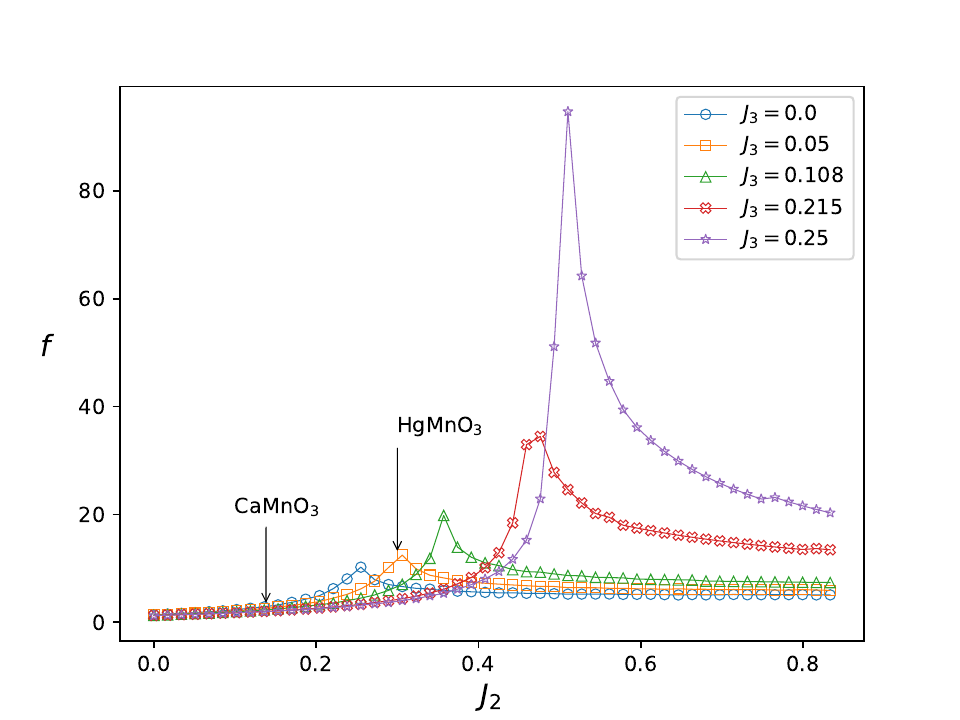}
  \caption{The ratio $f= |\theta|/T_{\tmop{N}}$ as a function of
    $J_2$ calculated in the Monte Carlo simulation for the same
    parameters as in Fig.~\ref{tn}\label{f-ratio} }
\end{figure*}

To quantify the degree of frustration, Fig.~\ref{f-ratio} presents for the same parameters as in Fig.~\ref{tn} the temperature dependence of the frustration ratio
\begin{eqnarray*}
  f  = |\theta|/T_{\tmop{N}}
\end{eqnarray*}
where  $\theta = -J({\bf q} = 0)/3$ is the Curie-Weiss (paramagnetic Curie) temperature. 
From this figure it is clear how the effects of the frustration
increase with increasing $J_3$.

\section{Conclusions}

To conclude, the Monte Carlo calculations convincingly demonstrate the importance of fluctuation effects. These effects considerably modify the results obtained from mean-field theory and the Tyablikov approximation. Therefore, accounting for fluctuations is crucial for the correct interpretation of experimental data on the thermodynamic properties of real antiferromagnetic compounds, particularly frustrated ones.  The inclusion of the third-neighbor exchange $J_3$ into the Hamiltonian increases the maximum of the frustration ratio $f$ and shifts its position
and the position of the Neel temperature's minimum.

The long-range antiferromagnetic exchange interaction with the second and third nearest neighbors becomes especially significant for $4d$ and $5d$ transition metal oxides. In these materials, the wave functions of $d$ electrons are more extended than in the $3d$ case due to the larger principal quantum number~\cite{streltsov2017}.

\section{Acknowledgment}
We thank E.~Komleva for the help with Fig.~\ref{model} and Y.~Iqbal for bringing
our attention to Ref. \cite{Balla_2019}. Computations were performed on the Uran
supercomputer at the IMM UB RAS.

\section{Funding}
The work was carried out within the framework of the state assignment of the Ministry of Science and Higher Education of the Russian Federation for the Mikheev Institute of Metal Physics, Ural Branch, Russian Academy of Sciences.

\section{Conflict of interests}
The authors of this work declare that they have no conflicts of interest.


\bibliography{bibliography}

\begin{thebibliography}{36}%
\makeatletter
\providecommand \@ifxundefined [1]{%
 \@ifx{#1\undefined}
}%
\providecommand \@ifnum [1]{%
 \ifnum #1\expandafter \@firstoftwo
 \else \expandafter \@secondoftwo
 \fi
}%
\providecommand \@ifx [1]{%
 \ifx #1\expandafter \@firstoftwo
 \else \expandafter \@secondoftwo
 \fi
}%
\providecommand \natexlab [1]{#1}%
\providecommand \enquote  [1]{``#1''}%
\providecommand \bibnamefont  [1]{#1}%
\providecommand \bibfnamefont [1]{#1}%
\providecommand \citenamefont [1]{#1}%
\providecommand \href@noop [0]{\@secondoftwo}%
\providecommand \href [0]{\begingroup \@sanitize@url \@href}%
\providecommand \@href[1]{\@@startlink{#1}\@@href}%
\providecommand \@@href[1]{\endgroup#1\@@endlink}%
\providecommand \@sanitize@url [0]{\catcode `\\12\catcode `\$12\catcode
  `\&12\catcode `\#12\catcode `\^12\catcode `\_12\catcode `\%12\relax}%
\providecommand \@@startlink[1]{}%
\providecommand \@@endlink[0]{}%
\providecommand \url  [0]{\begingroup\@sanitize@url \@url }%
\providecommand \@url [1]{\endgroup\@href {#1}{\urlprefix }}%
\providecommand \urlprefix  [0]{URL }%
\providecommand \Eprint [0]{\href }%
\providecommand \doibase [0]{https://doi.org/}%
\providecommand \selectlanguage [0]{\@gobble}%
\providecommand \bibinfo  [0]{\@secondoftwo}%
\providecommand \bibfield  [0]{\@secondoftwo}%
\providecommand \translation [1]{[#1]}%
\providecommand \BibitemOpen [0]{}%
\providecommand \bibitemStop [0]{}%
\providecommand \bibitemNoStop [0]{.\EOS\space}%
\providecommand \EOS [0]{\spacefactor3000\relax}%
\providecommand \BibitemShut  [1]{\csname bibitem#1\endcsname}%
\let\auto@bib@innerbib\@empty
\bibitem [{\citenamefont {Diep~(ed.)}(2004)}]{Diepbook}%
  \BibitemOpen
  \bibfield  {author} {\bibinfo {author} {\bibfnamefont {H.~T.}\ \bibnamefont
  {Diep~(ed.)}},\ }\href@noop {} {\emph {\bibinfo {title} {Frustrated Spin
  Systems}}}\ (\bibinfo  {publisher} {World Scientific Publishing Co. Pte.
  Ltd.},\ \bibinfo {year} {2004})\BibitemShut {NoStop}%
\bibitem [{\citenamefont {Khomskii}\ and\ \citenamefont
  {Streltsov}(2024)}]{khomskii2024}%
  \BibitemOpen
  \bibfield  {author} {\bibinfo {author} {\bibfnamefont {D.~I.}\ \bibnamefont
  {Khomskii}}\ and\ \bibinfo {author} {\bibfnamefont {S.~V.}\ \bibnamefont
  {Streltsov}},\ }\bibfield  {title} {\bibinfo {title} {{Magnetic Oxides}},\
  }in\ \href {https://doi.org/10.1016/B978-0-323-90800-9.00120-7} {\emph
  {\bibinfo {booktitle} {Encyclopedia of Condensed Matter Physics}}}\ (\bibinfo
   {publisher} {Elsevier},\ \bibinfo {year} {2024})\ pp.\ \bibinfo {pages}
  {98--111}\BibitemShut {NoStop}%
\bibitem [{\citenamefont {Balents}(2010)}]{Balents2010}%
  \BibitemOpen
  \bibfield  {author} {\bibinfo {author} {\bibfnamefont {L.}~\bibnamefont
  {Balents}},\ }\bibfield  {title} {\bibinfo {title} {{Spin liquids in
  frustrated magnets}},\ }\href {https://doi.org/10.1038/nature08917}
  {\bibfield  {journal} {\bibinfo  {journal} {Nature}\ }\textbf {\bibinfo
  {volume} {464}},\ \bibinfo {pages} {199} (\bibinfo {year}
  {2010})}\BibitemShut {NoStop}%
\bibitem [{\citenamefont {Gong}\ \emph {et~al.}(2013)\citenamefont {Gong},
  \citenamefont {Sheng}, \citenamefont {Motrunich},\ and\ \citenamefont
  {Fisher}}]{PhysRevB.88.165138}%
  \BibitemOpen
  \bibfield  {author} {\bibinfo {author} {\bibfnamefont {S.-S.}\ \bibnamefont
  {Gong}}, \bibinfo {author} {\bibfnamefont {D.~N.}\ \bibnamefont {Sheng}},
  \bibinfo {author} {\bibfnamefont {O.~I.}\ \bibnamefont {Motrunich}},\ and\
  \bibinfo {author} {\bibfnamefont {M.~P.~A.}\ \bibnamefont {Fisher}},\
  }\bibfield  {title} {\bibinfo {title} {{Phase diagram of the
  spin-$\frac{1}{2}$ ${J}_{1}$-${J}_{2}$ Heisenberg model on a honeycomb
  lattice}},\ }\href {https://doi.org/10.1103/PhysRevB.88.165138} {\bibfield
  {journal} {\bibinfo  {journal} {Phys. Rev. B}\ }\textbf {\bibinfo {volume}
  {88}},\ \bibinfo {pages} {165138} (\bibinfo {year} {2013})}\BibitemShut
  {NoStop}%
\bibitem [{\citenamefont {Iqbal}\ \emph {et~al.}(2016)\citenamefont {Iqbal},
  \citenamefont {Thomale}, \citenamefont {Parisen~Toldin}, \citenamefont
  {Rachel},\ and\ \citenamefont {Reuther}}]{PhysRevB.94.140408}%
  \BibitemOpen
  \bibfield  {author} {\bibinfo {author} {\bibfnamefont {Y.}~\bibnamefont
  {Iqbal}}, \bibinfo {author} {\bibfnamefont {R.}~\bibnamefont {Thomale}},
  \bibinfo {author} {\bibfnamefont {F.}~\bibnamefont {Parisen~Toldin}},
  \bibinfo {author} {\bibfnamefont {S.}~\bibnamefont {Rachel}},\ and\ \bibinfo
  {author} {\bibfnamefont {J.}~\bibnamefont {Reuther}},\ }\bibfield  {title}
  {\bibinfo {title} {{Functional renormalization group for three-dimensional
  quantum magnetism}},\ }\href {https://doi.org/10.1103/PhysRevB.94.140408}
  {\bibfield  {journal} {\bibinfo  {journal} {Phys. Rev. B}\ }\textbf {\bibinfo
  {volume} {94}},\ \bibinfo {pages} {140408} (\bibinfo {year}
  {2016})}\BibitemShut {NoStop}%
\bibitem [{\citenamefont {Oitmaa}(2017)}]{PhysRevB.95.014427}%
  \BibitemOpen
  \bibfield  {author} {\bibinfo {author} {\bibfnamefont {J.}~\bibnamefont
  {Oitmaa}},\ }\bibfield  {title} {\bibinfo {title} {{Frustrated
  ${J}_{1}\ensuremath{-}{J}_{2}\ensuremath{-}{J}_{3}$ Heisenberg
  antiferromagnet on the simple cubic lattice}},\ }\href
  {https://doi.org/10.1103/PhysRevB.95.014427} {\bibfield  {journal} {\bibinfo
  {journal} {Phys. Rev. B}\ }\textbf {\bibinfo {volume} {95}},\ \bibinfo
  {pages} {014427} (\bibinfo {year} {2017})}\BibitemShut {NoStop}%
\bibitem [{\citenamefont {Bri\'atico}\ \emph {et~al.}(1996)\citenamefont
  {Bri\'atico}, \citenamefont {Alascio}, \citenamefont {Allub}, \citenamefont
  {Butera}, \citenamefont {Caneiro}, \citenamefont {Causa},\ and\ \citenamefont
  {Tovar}}]{PhysRevB.53.14020}%
  \BibitemOpen
  \bibfield  {author} {\bibinfo {author} {\bibfnamefont {J.}~\bibnamefont
  {Bri\'atico}}, \bibinfo {author} {\bibfnamefont {B.}~\bibnamefont {Alascio}},
  \bibinfo {author} {\bibfnamefont {R.}~\bibnamefont {Allub}}, \bibinfo
  {author} {\bibfnamefont {A.}~\bibnamefont {Butera}}, \bibinfo {author}
  {\bibfnamefont {A.}~\bibnamefont {Caneiro}}, \bibinfo {author} {\bibfnamefont
  {M.~T.}\ \bibnamefont {Causa}},\ and\ \bibinfo {author} {\bibfnamefont
  {M.}~\bibnamefont {Tovar}},\ }\bibfield  {title} {\bibinfo {title}
  {{Double-exchange interaction in electron-doped
  $\mathrm{CaMn}{\mathrm{O}}_{3\ensuremath{-}\ensuremath{\delta}}$
  perovskites}},\ }\href {https://doi.org/10.1103/PhysRevB.53.14020} {\bibfield
   {journal} {\bibinfo  {journal} {Phys. Rev. B}\ }\textbf {\bibinfo {volume}
  {53}},\ \bibinfo {pages} {14020} (\bibinfo {year} {1996})}\BibitemShut
  {NoStop}%
\bibitem [{\citenamefont {Zhou}\ \emph {et~al.}(2020)\citenamefont {Zhou},
  \citenamefont {Qin}, \citenamefont {Ma}, \citenamefont {Ye}, \citenamefont
  {Guo}, \citenamefont {Yu}, \citenamefont {Lin}, \citenamefont {Chen},
  \citenamefont {Hu}, \citenamefont {Tjeng}, \citenamefont {Zhou},
  \citenamefont {Dong},\ and\ \citenamefont {Long}}]{Zhou2020}%
  \BibitemOpen
  \bibfield  {author} {\bibinfo {author} {\bibfnamefont {B.}~\bibnamefont
  {Zhou}}, \bibinfo {author} {\bibfnamefont {S.}~\bibnamefont {Qin}}, \bibinfo
  {author} {\bibfnamefont {T.}~\bibnamefont {Ma}}, \bibinfo {author}
  {\bibfnamefont {X.}~\bibnamefont {Ye}}, \bibinfo {author} {\bibfnamefont
  {J.}~\bibnamefont {Guo}}, \bibinfo {author} {\bibfnamefont {X.}~\bibnamefont
  {Yu}}, \bibinfo {author} {\bibfnamefont {H.-J.}\ \bibnamefont {Lin}},
  \bibinfo {author} {\bibfnamefont {C.-T.}\ \bibnamefont {Chen}}, \bibinfo
  {author} {\bibfnamefont {Z.}~\bibnamefont {Hu}}, \bibinfo {author}
  {\bibfnamefont {L.-H.}\ \bibnamefont {Tjeng}}, \bibinfo {author}
  {\bibfnamefont {G.}~\bibnamefont {Zhou}}, \bibinfo {author} {\bibfnamefont
  {C.}~\bibnamefont {Dong}},\ and\ \bibinfo {author} {\bibfnamefont
  {Y.}~\bibnamefont {Long}},\ }\bibfield  {title} {\bibinfo {title}
  {{High-Pressure Synthesis of Two Polymorphic HgMnO3 Phases and Distinct
  Magnetism from 2D to 3D}},\ }\href
  {https://doi.org/10.1021/acs.inorgchem.9b03551} {\bibfield  {journal}
  {\bibinfo  {journal} {Inorganic Chemistry}\ }\textbf {\bibinfo {volume}
  {59}},\ \bibinfo {pages} {3887} (\bibinfo {year} {2020})}\BibitemShut
  {NoStop}%
\bibitem [{\citenamefont {Myakotnikov}\ \emph {et~al.}(2024)\citenamefont
  {Myakotnikov}, \citenamefont {Komleva}, \citenamefont {Long},\ and\
  \citenamefont {Streltsov}}]{MyakotnikovPRB2024}%
  \BibitemOpen
  \bibfield  {author} {\bibinfo {author} {\bibfnamefont {D.~A.}\ \bibnamefont
  {Myakotnikov}}, \bibinfo {author} {\bibfnamefont {E.~V.}\ \bibnamefont
  {Komleva}}, \bibinfo {author} {\bibfnamefont {Y.}~\bibnamefont {Long}},\ and\
  \bibinfo {author} {\bibfnamefont {S.~V.}\ \bibnamefont {Streltsov}},\
  }\bibfield  {title} {\bibinfo {title} {{Importance of the indirect exchange
  interaction via $s$ states in altermagnetic ${\mathrm{HgMnO}}_{3}$}},\ }\href
  {https://doi.org/10.1103/PhysRevB.110.134427} {\bibfield  {journal} {\bibinfo
   {journal} {Phys. Rev. B}\ }\textbf {\bibinfo {volume} {110}},\ \bibinfo
  {pages} {134427} (\bibinfo {year} {2024})}\BibitemShut {NoStop}%
\bibitem [{\citenamefont {Tarasevych}\ \emph {et~al.}(2022)\citenamefont
  {Tarasevych}, \citenamefont {R\"uckriegel}, \citenamefont {Keupert},
  \citenamefont {Mitsiioannou},\ and\ \citenamefont
  {Kopietz}}]{PhysRevB.106.174412}%
  \BibitemOpen
  \bibfield  {author} {\bibinfo {author} {\bibfnamefont {D.}~\bibnamefont
  {Tarasevych}}, \bibinfo {author} {\bibfnamefont {A.}~\bibnamefont
  {R\"uckriegel}}, \bibinfo {author} {\bibfnamefont {S.}~\bibnamefont
  {Keupert}}, \bibinfo {author} {\bibfnamefont {V.}~\bibnamefont
  {Mitsiioannou}},\ and\ \bibinfo {author} {\bibfnamefont {P.}~\bibnamefont
  {Kopietz}},\ }\bibfield  {title} {\bibinfo {title} {{Spin-functional
  renormalization group for the ${J}_{1}{J}_{2}{J}_{3}$ quantum Heisenberg
  model}},\ }\href {https://doi.org/10.1103/PhysRevB.106.174412} {\bibfield
  {journal} {\bibinfo  {journal} {Phys. Rev. B}\ }\textbf {\bibinfo {volume}
  {106}},\ \bibinfo {pages} {174412} (\bibinfo {year} {2022})}\BibitemShut
  {NoStop}%
\bibitem [{\citenamefont {Hu}\ and\ \citenamefont {Wang}(2019)}]{HuWang2019}%
  \BibitemOpen
  \bibfield  {author} {\bibinfo {author} {\bibfnamefont {A.-Y.}\ \bibnamefont
  {Hu}}\ and\ \bibinfo {author} {\bibfnamefont {H.-Y.}\ \bibnamefont {Wang}},\
  }\bibfield  {title} {\bibinfo {title} {{Phase transition of the frustrated
  antiferromagntic J1-J2-J3 spin-1/2 Heisenberg model on a simple cubic
  lattice}},\ }\href
  {https://doi.org/https://doi.org/10.1007/s11467-018-0831-x} {\bibfield
  {journal} {\bibinfo  {journal} {Frontiers of Physics}\ }\textbf {\bibinfo
  {volume} {14}},\ \bibinfo {pages} {13605} (\bibinfo {year}
  {2019})}\BibitemShut {NoStop}%
\bibitem [{\citenamefont {Irkhin}\ \emph {et~al.}(1992)\citenamefont {Irkhin},
  \citenamefont {Katanin},\ and\ \citenamefont {Katsnelson}}]{VYuIrkhin1992}%
  \BibitemOpen
  \bibfield  {author} {\bibinfo {author} {\bibfnamefont {V.~Y.}\ \bibnamefont
  {Irkhin}}, \bibinfo {author} {\bibfnamefont {A.~A.}\ \bibnamefont
  {Katanin}},\ and\ \bibinfo {author} {\bibfnamefont {M.~I.}\ \bibnamefont
  {Katsnelson}},\ }\bibfield  {title} {\bibinfo {title} {{On the
  self-consistent spin-wave theory of frustrated Heisenberg
  antiferromagnets}},\ }\href {https://doi.org/10.1088/0953-8984/4/22/019}
  {\bibfield  {journal} {\bibinfo  {journal} {Journal of Physics: Condensed
  Matter}\ }\textbf {\bibinfo {volume} {4}},\ \bibinfo {pages} {5227} (\bibinfo
  {year} {1992})}\BibitemShut {NoStop}%
\bibitem [{\citenamefont {Majumdar}\ and\ \citenamefont
  {Datta}(2010)}]{MajumdarDatta}%
  \BibitemOpen
  \bibfield  {author} {\bibinfo {author} {\bibfnamefont {K.}~\bibnamefont
  {Majumdar}}\ and\ \bibinfo {author} {\bibfnamefont {T.}~\bibnamefont
  {Datta}},\ }\bibfield  {title} {\bibinfo {title} {{Zero Temperature Phases of
  the Frustrated J1-J2 Antiferromagnetic Spin-1/2 Heisenberg Model on a Simple
  Cubic Lattice}},\ }\href {https://doi.org/10.1007/s10955-010-9967-y}
  {\bibfield  {journal} {\bibinfo  {journal} {Journal of Statistical Physics}\
  ,\ \bibinfo {pages} {714–726}} (\bibinfo {year} {2010})}\BibitemShut
  {NoStop}%
\bibitem [{\citenamefont {Sandvik}(1998)}]{PhysRevLett.80.5196}%
  \BibitemOpen
  \bibfield  {author} {\bibinfo {author} {\bibfnamefont {A.~W.}\ \bibnamefont
  {Sandvik}},\ }\bibfield  {title} {\bibinfo {title} {{Critical Temperature and
  the Transition from Quantum to Classical Order Parameter Fluctuations in the
  Three-Dimensional Heisenberg Antiferromagnet}},\ }\href
  {https://doi.org/10.1103/PhysRevLett.80.5196} {\bibfield  {journal} {\bibinfo
   {journal} {Phys. Rev. Lett.}\ }\textbf {\bibinfo {volume} {80}},\ \bibinfo
  {pages} {5196} (\bibinfo {year} {1998})}\BibitemShut {NoStop}%
\bibitem [{\citenamefont {Pinettes}\ and\ \citenamefont
  {Diep}(1998)}]{Pinettes1998}%
  \BibitemOpen
  \bibfield  {author} {\bibinfo {author} {\bibfnamefont {C.}~\bibnamefont
  {Pinettes}}\ and\ \bibinfo {author} {\bibfnamefont {H.~T.}\ \bibnamefont
  {Diep}},\ }\bibfield  {title} {\bibinfo {title} {{Phase transition and phase
  diagram of the J1-J2 Heisenberg model on a simple cubic lattice}},\ }\href
  {https://doi.org/10.1063/1.367729} {\bibfield  {journal} {\bibinfo  {journal}
  {Journal of Applied Physics}\ }\textbf {\bibinfo {volume} {83}},\ \bibinfo
  {pages} {6317} (\bibinfo {year} {1998})}\BibitemShut {NoStop}%
\bibitem [{\citenamefont {Ngo}\ \emph {et~al.}(2011)\citenamefont {Ngo},
  \citenamefont {Hoang},\ and\ \citenamefont
  {Diep}}]{doi:10.1142/S0217984911026632}%
  \BibitemOpen
  \bibfield  {author} {\bibinfo {author} {\bibfnamefont {V.~T.}\ \bibnamefont
  {Ngo}}, \bibinfo {author} {\bibfnamefont {D.~T.}\ \bibnamefont {Hoang}},\
  and\ \bibinfo {author} {\bibfnamefont {H.~T.}\ \bibnamefont {Diep}},\
  }\bibfield  {title} {\bibinfo {title} {{Phase transition in the Heisenberg
  fully-frustrated simple cubic lattice}},\ }\href
  {https://doi.org/10.1142/S0217984911026632} {\bibfield  {journal} {\bibinfo
  {journal} {Modern Physics Letters B}\ }\textbf {\bibinfo {volume} {25}},\
  \bibinfo {pages} {929} (\bibinfo {year} {2011})},\ \Eprint
  {https://arxiv.org/abs/https://doi.org/10.1142/S0217984911026632}
  {https://doi.org/10.1142/S0217984911026632} \BibitemShut {NoStop}%
\bibitem [{\citenamefont {Murtazaev}\ \emph {et~al.}(2020)\citenamefont
  {Murtazaev}, \citenamefont {Kassan-Ogly}, \citenamefont {Ramazanov},\ and\
  \citenamefont {Murtazaev}}]{Murtazaev2020}%
  \BibitemOpen
  \bibfield  {author} {\bibinfo {author} {\bibfnamefont {A.~K.}\ \bibnamefont
  {Murtazaev}}, \bibinfo {author} {\bibfnamefont {F.~A.}\ \bibnamefont
  {Kassan-Ogly}}, \bibinfo {author} {\bibfnamefont {M.~K.}\ \bibnamefont
  {Ramazanov}},\ and\ \bibinfo {author} {\bibfnamefont {K.~S.}\ \bibnamefont
  {Murtazaev}},\ }\bibfield  {title} {\bibinfo {title} {{Study of Phase
  Transitions in the Antiferromagnetic Heisenberg Model on a Body-Centered
  Cubic Lattice by Monte Carlo Simulation}},\ }\href
  {https://doi.org/10.1134/S0031918X20040109} {\bibfield  {journal} {\bibinfo
  {journal} {Physics of Metals and Metallography}\ }\textbf {\bibinfo {volume}
  {121}},\ \bibinfo {pages} {305} (\bibinfo {year} {2020})}\BibitemShut
  {NoStop}%
\bibitem [{\citenamefont {Sharafullin}\ \emph {et~al.}(2023)\citenamefont
  {Sharafullin}, \citenamefont {Yuldasheva}, \citenamefont {Abdrakhmanov},\
  and\ \citenamefont {Nugumanov}}]{Sharafullin2023}%
  \BibitemOpen
  \bibfield  {author} {\bibinfo {author} {\bibfnamefont {I.~F.}\ \bibnamefont
  {Sharafullin}}, \bibinfo {author} {\bibfnamefont {A.~R.}\ \bibnamefont
  {Yuldasheva}}, \bibinfo {author} {\bibfnamefont {D.~I.}\ \bibnamefont
  {Abdrakhmanov}},\ and\ \bibinfo {author} {\bibfnamefont {A.~G.}\ \bibnamefont
  {Nugumanov}},\ }\bibfield  {title} {\bibinfo {title} {{Skyrmion Lattices
  Phase Driven by Interfacial-Engineered Dzyaloshinskii--Moriya Interaction in
  Frustrated Antiferromagnetic/Ferroelectric Bilayers}},\ }\href
  {https://doi.org/10.1134/S0031918X23601452} {\bibfield  {journal} {\bibinfo
  {journal} {Physics of Metals and Metallography}\ }\textbf {\bibinfo {volume}
  {124}},\ \bibinfo {pages} {1697} (\bibinfo {year} {2023})}\BibitemShut
  {NoStop}%
\bibitem [{\citenamefont {Zhou}\ \emph {et~al.}(2023)\citenamefont {Zhou},
  \citenamefont {Chen},\ and\ \citenamefont {Li}}]{ZepengZhou2023}%
  \BibitemOpen
  \bibfield  {author} {\bibinfo {author} {\bibfnamefont {Z.}~\bibnamefont
  {Zhou}}, \bibinfo {author} {\bibfnamefont {Y.}~\bibnamefont {Chen}},\ and\
  \bibinfo {author} {\bibfnamefont {W.}~\bibnamefont {Li}},\ }\bibfield
  {title} {\bibinfo {title} {{Onsager Reaction Field Theory for Two-Dimensional
  Spatially Anisotropic Heisenberg Ferromagnet with the x-Axis Long-Range
  Interaction}},\ }\href {https://doi.org/10.1134/S0031918X23600744} {\bibfield
   {journal} {\bibinfo  {journal} {Physics of Metals and Metallography}\
  }\textbf {\bibinfo {volume} {124}},\ \bibinfo {pages} {1716} (\bibinfo {year}
  {2023})}\BibitemShut {NoStop}%
\bibitem [{\citenamefont {Kolesnikov}\ \emph {et~al.}(2024)\citenamefont
  {Kolesnikov}, \citenamefont {Sapronova},\ and\ \citenamefont
  {Saletsky}}]{Kolesnikov2024}%
  \BibitemOpen
  \bibfield  {author} {\bibinfo {author} {\bibfnamefont {S.~V.}\ \bibnamefont
  {Kolesnikov}}, \bibinfo {author} {\bibfnamefont {E.~S.}\ \bibnamefont
  {Sapronova}},\ and\ \bibinfo {author} {\bibfnamefont {A.~M.}\ \bibnamefont
  {Saletsky}},\ }\bibfield  {title} {\bibinfo {title} {{Remagnetization of
  Finite-Length Ferromagnetic Cobalt Atomic Chains}},\ }\href
  {https://doi.org/10.1134/S0031918X2460057X} {\bibfield  {journal} {\bibinfo
  {journal} {Physics of Metals and Metallography}\ }\textbf {\bibinfo {volume}
  {125}},\ \bibinfo {pages} {683} (\bibinfo {year} {2024})}\BibitemShut
  {NoStop}%
\bibitem [{\citenamefont {Ignatenko}\ and\ \citenamefont
  {Irkhin}(2016)}]{Ignatenko2016}%
  \BibitemOpen
  \bibfield  {author} {\bibinfo {author} {\bibfnamefont {A.~N.}\ \bibnamefont
  {Ignatenko}}\ and\ \bibinfo {author} {\bibfnamefont {V.~Y.}\ \bibnamefont
  {Irkhin}},\ }\bibfield  {title} {\bibinfo {title} {{Frustrated Heisenberg
  Antiferromagnets on Cubic Lattices: Magnetic Structures, Exchange Gaps, and
  Non-Conventional Critical Behaviour}},\ }\href@noop {} {\bibfield  {journal}
  {\bibinfo  {journal} {Journal of Siberian Federal University. Mathematics \&
  Physics}\ }\textbf {\bibinfo {volume} {9}},\ \bibinfo {pages} {454} (\bibinfo
  {year} {2016})}\BibitemShut {NoStop}%
\bibitem [{\citenamefont {Gaenko}\ \emph {et~al.}(2017)\citenamefont {Gaenko},
  \citenamefont {Antipov}, \citenamefont {Carcassi}, \citenamefont {Chen},
  \citenamefont {Chen}, \citenamefont {Dong}, \citenamefont {Gamper},
  \citenamefont {Gukelberger}, \citenamefont {Igarashi}, \citenamefont
  {Iskakov}, \citenamefont {Könz}, \citenamefont {LeBlanc}, \citenamefont
  {Levy}, \citenamefont {Ma}, \citenamefont {Paki}, \citenamefont {Shinaoka},
  \citenamefont {Todo}, \citenamefont {Troyer},\ and\ \citenamefont
  {Gull}}]{GAENKO2017235}%
  \BibitemOpen
  \bibfield  {author} {\bibinfo {author} {\bibfnamefont {A.}~\bibnamefont
  {Gaenko}}, \bibinfo {author} {\bibfnamefont {A.}~\bibnamefont {Antipov}},
  \bibinfo {author} {\bibfnamefont {G.}~\bibnamefont {Carcassi}}, \bibinfo
  {author} {\bibfnamefont {T.}~\bibnamefont {Chen}}, \bibinfo {author}
  {\bibfnamefont {X.}~\bibnamefont {Chen}}, \bibinfo {author} {\bibfnamefont
  {Q.}~\bibnamefont {Dong}}, \bibinfo {author} {\bibfnamefont {L.}~\bibnamefont
  {Gamper}}, \bibinfo {author} {\bibfnamefont {J.}~\bibnamefont {Gukelberger}},
  \bibinfo {author} {\bibfnamefont {R.}~\bibnamefont {Igarashi}}, \bibinfo
  {author} {\bibfnamefont {S.}~\bibnamefont {Iskakov}}, \bibinfo {author}
  {\bibfnamefont {M.}~\bibnamefont {Könz}}, \bibinfo {author} {\bibfnamefont
  {J.}~\bibnamefont {LeBlanc}}, \bibinfo {author} {\bibfnamefont
  {R.}~\bibnamefont {Levy}}, \bibinfo {author} {\bibfnamefont {P.}~\bibnamefont
  {Ma}}, \bibinfo {author} {\bibfnamefont {J.}~\bibnamefont {Paki}}, \bibinfo
  {author} {\bibfnamefont {H.}~\bibnamefont {Shinaoka}}, \bibinfo {author}
  {\bibfnamefont {S.}~\bibnamefont {Todo}}, \bibinfo {author} {\bibfnamefont
  {M.}~\bibnamefont {Troyer}},\ and\ \bibinfo {author} {\bibfnamefont
  {E.}~\bibnamefont {Gull}},\ }\bibfield  {title} {\bibinfo {title} {{Updated
  core libraries of the ALPS project}},\ }\href
  {https://doi.org/https://doi.org/10.1016/j.cpc.2016.12.009} {\bibfield
  {journal} {\bibinfo  {journal} {Computer Physics Communications}\ }\textbf
  {\bibinfo {volume} {213}},\ \bibinfo {pages} {235} (\bibinfo {year}
  {2017})}\BibitemShut {NoStop}%
\bibitem [{\citenamefont {Vasiliev}\ \emph {et~al.}(2019)\citenamefont
  {Vasiliev}, \citenamefont {Volkova}, \citenamefont {Zvereva},\ and\
  \citenamefont {Markina}}]{vasiliev-book}%
  \BibitemOpen
  \bibfield  {author} {\bibinfo {author} {\bibfnamefont {A.~N.}\ \bibnamefont
  {Vasiliev}}, \bibinfo {author} {\bibfnamefont {O.~S.}\ \bibnamefont
  {Volkova}}, \bibinfo {author} {\bibfnamefont {E.~A.}\ \bibnamefont
  {Zvereva}},\ and\ \bibinfo {author} {\bibfnamefont {M.~M.}\ \bibnamefont
  {Markina}},\ }\href@noop {} {\emph {\bibinfo {title} {{Low-Dimensional
  Magnetism}}}}\ (\bibinfo  {publisher} {CRC Press},\ \bibinfo {year}
  {2019})\BibitemShut {NoStop}%
\bibitem [{\citenamefont {Shender}(1982)}]{Shender1982}%
  \BibitemOpen
  \bibfield  {author} {\bibinfo {author} {\bibfnamefont {E.}~\bibnamefont
  {Shender}},\ }\bibfield  {title} {\bibinfo {title} {{Antiferromagnetic
  garnets with fluctuationally interacting sublattices}},\ }\href
  {http://www.jetp.ac.ru/cgi-bin/r/index/e/56/1/p178?a=list} {\bibfield
  {journal} {\bibinfo  {journal} {JETP}\ }\textbf {\bibinfo {volume} {56}},\
  \bibinfo {pages} {178} (\bibinfo {year} {1982})}\BibitemShut {NoStop}%
\bibitem [{\citenamefont {Yildirim}\ \emph {et~al.}(1998)\citenamefont
  {Yildirim}, \citenamefont {Harris},\ and\ \citenamefont
  {Shender}}]{PhysRevB.58.3144}%
  \BibitemOpen
  \bibfield  {author} {\bibinfo {author} {\bibfnamefont {T.}~\bibnamefont
  {Yildirim}}, \bibinfo {author} {\bibfnamefont {A.~B.}\ \bibnamefont
  {Harris}},\ and\ \bibinfo {author} {\bibfnamefont {E.~F.}\ \bibnamefont
  {Shender}},\ }\bibfield  {title} {\bibinfo {title} {{Frustration and quantum
  fluctuations in Heisenberg fcc antiferromagnets}},\ }\href
  {https://doi.org/10.1103/PhysRevB.58.3144} {\bibfield  {journal} {\bibinfo
  {journal} {Phys. Rev. B}\ }\textbf {\bibinfo {volume} {58}},\ \bibinfo
  {pages} {3144} (\bibinfo {year} {1998})}\BibitemShut {NoStop}%
\bibitem [{\citenamefont {Ignatenko}\ and\ \citenamefont
  {Irkhin}(2023)}]{Ignatenko2023}%
  \BibitemOpen
  \bibfield  {author} {\bibinfo {author} {\bibfnamefont {A.~N.}\ \bibnamefont
  {Ignatenko}}\ and\ \bibinfo {author} {\bibfnamefont {V.~Y.}\ \bibnamefont
  {Irkhin}},\ }\bibfield  {title} {\bibinfo {title} {{The Ising Nematic in the
  J1--J2 Heisenberg Model on a Square Lattice in a Self-Consistent Spin-Wave
  Theory}},\ }\href {https://doi.org/10.3103/S1062873823703768} {\bibfield
  {journal} {\bibinfo  {journal} {Bulletin of the Russian Academy of Sciences:
  Physics}\ }\textbf {\bibinfo {volume} {87}},\ \bibinfo {pages} {1601}
  (\bibinfo {year} {2023})}\BibitemShut {NoStop}%
\bibitem [{\citenamefont {Ignatenko}\ \emph {et~al.}(2013)\citenamefont
  {Ignatenko}, \citenamefont {Katanin},\ and\ \citenamefont
  {Irkhin}}]{Ignatenko2013}%
  \BibitemOpen
  \bibfield  {author} {\bibinfo {author} {\bibfnamefont {A.~N.}\ \bibnamefont
  {Ignatenko}}, \bibinfo {author} {\bibfnamefont {A.~A.}\ \bibnamefont
  {Katanin}},\ and\ \bibinfo {author} {\bibfnamefont {V.~Y.}\ \bibnamefont
  {Irkhin}},\ }\bibfield  {title} {\bibinfo {title} {{Strong fluctuations near
  the frustration point in cubic lattice ferromagnets with localized
  moments}},\ }\href {https://doi.org/10.1134/S0021364013040073} {\bibfield
  {journal} {\bibinfo  {journal} {JETP Letters}\ }\textbf {\bibinfo {volume}
  {97}},\ \bibinfo {pages} {209} (\bibinfo {year} {2013})}\BibitemShut
  {NoStop}%
\bibitem [{\citenamefont {Privman~(ed.)}(1990)}]{Privman_book_1990}%
  \BibitemOpen
  \bibfield  {author} {\bibinfo {author} {\bibfnamefont {V.}~\bibnamefont
  {Privman~(ed.)}},\ }\bibinfo {title} {{Finite Size Scaling and Numerical
  Simulation of Statistical Systems}}\ (\bibinfo  {publisher} {World
  Scientific},\ \bibinfo {year} {1990})\ \Eprint
  {https://arxiv.org/abs/https://www.worldscientific.com/doi/pdf/10.1142/1011}
  {https://www.worldscientific.com/doi/pdf/10.1142/1011} \BibitemShut {NoStop}%
\bibitem [{\citenamefont {Binder}(1981)}]{Binder_PRL_1981}%
  \BibitemOpen
  \bibfield  {author} {\bibinfo {author} {\bibfnamefont {K.}~\bibnamefont
  {Binder}},\ }\bibfield  {title} {\bibinfo {title} {{Critical Properties from
  Monte Carlo Coarse Graining and Renormalization}},\ }\href
  {https://doi.org/10.1103/PhysRevLett.47.693} {\bibfield  {journal} {\bibinfo
  {journal} {Phys. Rev. Lett.}\ }\textbf {\bibinfo {volume} {47}},\ \bibinfo
  {pages} {693} (\bibinfo {year} {1981})}\BibitemShut {NoStop}%
\bibitem [{\citenamefont {Binder}\ and\ \citenamefont
  {Landau}(1984)}]{Binder_PRB_1984}%
  \BibitemOpen
  \bibfield  {author} {\bibinfo {author} {\bibfnamefont {K.}~\bibnamefont
  {Binder}}\ and\ \bibinfo {author} {\bibfnamefont {D.~P.}\ \bibnamefont
  {Landau}},\ }\bibfield  {title} {\bibinfo {title} {{Finite-size scaling at
  first-order phase transitions}},\ }\href
  {https://doi.org/10.1103/PhysRevB.30.1477} {\bibfield  {journal} {\bibinfo
  {journal} {Phys. Rev. B}\ }\textbf {\bibinfo {volume} {30}},\ \bibinfo
  {pages} {1477} (\bibinfo {year} {1984})}\BibitemShut {NoStop}%
\bibitem [{\citenamefont {Vollmayr}\ \emph {et~al.}(1993)\citenamefont
  {Vollmayr}, \citenamefont {Reger}, \citenamefont {Scheucher},\ and\
  \citenamefont {Binder}}]{Vollmayr1993}%
  \BibitemOpen
  \bibfield  {author} {\bibinfo {author} {\bibfnamefont {K.}~\bibnamefont
  {Vollmayr}}, \bibinfo {author} {\bibfnamefont {J.~D.}\ \bibnamefont {Reger}},
  \bibinfo {author} {\bibfnamefont {M.}~\bibnamefont {Scheucher}},\ and\
  \bibinfo {author} {\bibfnamefont {K.}~\bibnamefont {Binder}},\ }\bibfield
  {title} {\bibinfo {title} {{Finite size effects at thermally-driven first
  order phase transitions: A phenomenological theory of the order parameter
  distribution}},\ }\href {https://doi.org/10.1007/BF01316713} {\bibfield
  {journal} {\bibinfo  {journal} {Zeitschrift f{\"u}r Physik B Condensed
  Matter}\ }\textbf {\bibinfo {volume} {91}},\ \bibinfo {pages} {113} (\bibinfo
  {year} {1993})}\BibitemShut {NoStop}%
\bibitem [{\citenamefont {Tsai}\ and\ \citenamefont
  {Salinas}(1998)}]{TsaiSalinas1998}%
  \BibitemOpen
  \bibfield  {author} {\bibinfo {author} {\bibfnamefont {S.-H.}\ \bibnamefont
  {Tsai}}\ and\ \bibinfo {author} {\bibfnamefont {S.~R.}\ \bibnamefont
  {Salinas}},\ }\bibfield  {title} {\bibinfo {title} {{Fourth-Order Cumulants
  to Characterize the Phase Transitions of a Spin-1 Ising Model}},\ }\href
  {https://doi.org/10.1590/S0103-97331998000100008} {\bibfield  {journal}
  {\bibinfo  {journal} {Brazilian Journal of Physics}\ }\textbf {\bibinfo
  {volume} {28}},\ \bibinfo {pages} {58} (\bibinfo {year} {1998})}\BibitemShut
  {NoStop}%
\bibitem [{\citenamefont {Campostrini}\ \emph {et~al.}(2002)\citenamefont
  {Campostrini}, \citenamefont {Hasenbusch}, \citenamefont {Pelissetto},
  \citenamefont {Rossi},\ and\ \citenamefont {Vicari}}]{Campostrini_PRB_2002}%
  \BibitemOpen
  \bibfield  {author} {\bibinfo {author} {\bibfnamefont {M.}~\bibnamefont
  {Campostrini}}, \bibinfo {author} {\bibfnamefont {M.}~\bibnamefont
  {Hasenbusch}}, \bibinfo {author} {\bibfnamefont {A.}~\bibnamefont
  {Pelissetto}}, \bibinfo {author} {\bibfnamefont {P.}~\bibnamefont {Rossi}},\
  and\ \bibinfo {author} {\bibfnamefont {E.}~\bibnamefont {Vicari}},\
  }\bibfield  {title} {\bibinfo {title} {{Critical exponents and equation of
  state of the three-dimensional Heisenberg universality class}},\ }\href
  {https://doi.org/10.1103/PhysRevB.65.144520} {\bibfield  {journal} {\bibinfo
  {journal} {Phys. Rev. B}\ }\textbf {\bibinfo {volume} {65}},\ \bibinfo
  {pages} {144520} (\bibinfo {year} {2002})}\BibitemShut {NoStop}%
\bibitem [{\citenamefont {Brazovskii}\ \emph {et~al.}(1976)\citenamefont
  {Brazovskii}, \citenamefont {Dzyaloshinskii},\ and\ \citenamefont
  {Kukharenko}}]{Brazovskii1976}%
  \BibitemOpen
  \bibfield  {author} {\bibinfo {author} {\bibfnamefont {S.}~\bibnamefont
  {Brazovskii}}, \bibinfo {author} {\bibfnamefont {I.}~\bibnamefont
  {Dzyaloshinskii}},\ and\ \bibinfo {author} {\bibfnamefont {B.}~\bibnamefont
  {Kukharenko}},\ }\bibfield  {title} {\bibinfo {title} {{First-order magnetic
  phase transitions and fluctuations}},\ }\href
  {http://jetp.ras.ru/cgi-bin/dn/e_043_06_1178.pdf} {\bibfield  {journal}
  {\bibinfo  {journal} {Journal of Experimental and Theoretical Physics}\
  }\textbf {\bibinfo {volume} {43}},\ \bibinfo {pages} {1178} (\bibinfo {year}
  {1976})}\BibitemShut {NoStop}%
\bibitem [{\citenamefont {Balla}\ \emph {et~al.}(2019)\citenamefont {Balla},
  \citenamefont {Iqbal},\ and\ \citenamefont {Penc}}]{Balla_2019}%
  \BibitemOpen
  \bibfield  {author} {\bibinfo {author} {\bibfnamefont {P.}~\bibnamefont
  {Balla}}, \bibinfo {author} {\bibfnamefont {Y.}~\bibnamefont {Iqbal}},\ and\
  \bibinfo {author} {\bibfnamefont {K.}~\bibnamefont {Penc}},\ }\bibfield
  {title} {\bibinfo {title} {Affine lattice construction of spiral surfaces in
  frustrated heisenberg models},\ }\href
  {https://doi.org/10.1103/PhysRevB.100.140402} {\bibfield  {journal} {\bibinfo
   {journal} {Phys. Rev. B}\ }\textbf {\bibinfo {volume} {100}},\ \bibinfo
  {pages} {140402} (\bibinfo {year} {2019})}\BibitemShut {NoStop}%
\bibitem [{\citenamefont {Streltsov}\ and\ \citenamefont
  {Khomskii}(2017)}]{streltsov2017}%
  \BibitemOpen
  \bibfield  {author} {\bibinfo {author} {\bibfnamefont {S.}~\bibnamefont
  {Streltsov}}\ and\ \bibinfo {author} {\bibfnamefont {D.}~\bibnamefont
  {Khomskii}},\ }\bibfield  {title} {\bibinfo {title} {{Orbital Physics in
  Transition Metal Compounds: New Trends}},\ }\href
  {https://doi.org/10.3367/UFNe.2017.08.038196} {\bibfield  {journal} {\bibinfo
   {journal} {Physics-Uspekhi}\ }\textbf {\bibinfo {volume} {60}},\ \bibinfo
  {pages} {1121} (\bibinfo {year} {2017})}\BibitemShut {NoStop}%
\end{thebibliography}%

\end{document}